\documentclass[twocolumn,showpacs,showkeys]{revtex4}
\usepackage{graphicx}
\usepackage{bm}
\usepackage{color}
\usepackage{amsmath}
\usepackage{natbib}

\begin{document}

\title{Quantum hydrodynamic theory of quantum fluctuations in dipolar Bose-Einstein condensate}

\author{Pavel A. Andreev}
\email{andreevpa@physics.msu.ru}
\affiliation{Faculty of physics, Lomonosov Moscow State University, Moscow, Russian Federation, 119991.}

\date{\today}

\begin{abstract}
Traditional quantum hydrodynamics of Bose-Einstein condensates (BECs) is restricted by the continuity and Euler equations.
It corresponds to the well-known Gross-Pitaevskii equation.
However, the quantum Bohm potential, which is a part of the momentum flux, has a nontrivial part with can evolve under the quantum fluctuations.
To cover this phenomenon in terms of hydrodynamic methods
we need to derive equations for the second rank tensor (the momentum flux), and the third rank tensor.
In atomic BECs the interaction is the snort-range interaction.
In all equations we consider the main contribution of the short-range interaction
which appears in the first order by the interaction radius.
Derived hydrodynamics consists of four hydrodynamic equations.
However, two equations contain interaction.
The Euler equation contains interaction in the Gross-Pitaevskii approximation.
The third moment evolution equation contains interaction leading to the quantum fluctuations.
It is proportional to new interaction constant.
The Gross-Pitaevskii interaction constant is the integral of potential, but
the second interaction constant is the integral of second derivative of potential.
If we have dipolar BECs we deal with a long-range interaction.
Its contribution is proportional to the potential of dipole-dipole interaction (DDI) in the mean field regime.
The Euler equation contains the derivative of the potential.
The third rank tensor evolution equation contains the third derivative of the potential
which is also proportional to the square of the Plank constant.
It is responsible for the dipolar part of quantum fluctuations.
Higher derivatives correspond to the small scale contributions of the DDI.
The quantum  fluctuations lead to existence of the second wave solution.
Moreover, the quantum  fluctuations introduce the instability of the BECs.
If the dipole-dipole interaction is attractive,
but being smaller then the repulsive SRI presented by the first interaction constant,
there is the long-wavelength instability.
This scenario can be realized for dysprosium at $a=70a_{B}$,
where $a_{B}$ is the Borh radius.
For the repulsive DDI these is more complex picture.
There is the small area with the long-wavelength instability
which transits into stability interval,
where two waves exist.
And there is the short-wavelength instability,
which is stronger then the long-wavelength instability.
These results are found for the DDI strength comparable with the Gross-Pitaevskii short-range interaction,
while the dimensionless second interaction constant is $10\div100$ times smaller then the Gross-Pitaevskii interaction constant.
\end{abstract}



\maketitle


Hydrodynamics is a method of description of classic and quantum phenomena.
Mostly, it is applied to collective phenomena,
but it can be applied to the single quantum particle \cite{Madelung ZP 26}, \cite{Takabayasi PTP 54}, \cite{Takabayasi PTP 55 a}, \cite{Takabayasi PTP 83}.
Different forms of the force fields distinguish the hydrodynamics of different physical systems.
The Navier-–Stokes momentum equation describes the water flows and classic liquids \emph{and}
classic gases and some atmospheric phenomena \cite{Wyngaard ARFM 92}.
Hydrodynamics with the Euler equation containing the Lorentz force describes the waves,
instabilities and other collective phenomena in plasmas \cite{Gomberoff PRE 97}, \cite{Thompson PP 12}.
The quantum Bohm potential and spin-effects give an extension of hydrodynamics suitable for the quantum plasmas
\cite{Mahajan PRL 11}, \cite{Koide PRC 13}, \cite{Andreev EPL 16}, \cite{Andreev APL 16}.
The Gross-Pitaevskii equation for the Bose-Einstein condensate (BEC) of neutral atoms
can be presented as the set of two hydrodynamic equations \cite{Dalfovo RMP 99}, \cite{Fetter RMP 09}.
Spinor BECs can be also modeled by corresponding hydrodynamics
\cite{Szirmai PRA 12}, \cite{Stamper-Kurn RMP 13}, \cite{Fujimoto PRA 13}.


The described examples are based on different hydrodynamics,
which have same feature.
They are the composition of two equations: the continuity equation and the Euler equation.
The spin introduces additional equations, but our conclusion is about spinless part of dynamics.

There are examples of extended hydrodynamics,
as an example mention the classic plasmas \cite{Miller PoP 16},
where higher moments of the distribution function are adopted,
including the second order tensor evolution.

However, if we want to capture the quantum fluctuations in BECs purely from hydrodynamics
we need to derive two additional hydrodynamic equations.
They are the pressure $p^{\alpha\beta}$ evolution equation,
which is the second rank tensor existing in the Euler equation,
and the third rank tensor $Q^{\alpha\beta\gamma}$ evolution equation.

The BECs is the collection of bosons being in the lowest energy level.
Hence, we expect that the pressure $p^{\alpha\beta}$ and the third rank tensor $Q^{\alpha\beta\gamma}$ are equal to zero
since their definitions explicitly refer to the presence of particles in excited states.
However, these tensors have a source in the quantum theory.
It is located in equation for tensor $Q^{\alpha\beta\gamma}$
which proportional to square of the Planck constant $\hbar^{2}$ and the interaction potential $g_{2}\sim\int d\textbf{r} U''(r)$
(for the short-range interaction).
Corresponding model is presented in this paper for the dipolar BECs.
Therefore, it covers specifics related to the short-range interaction
and the long-range interaction presented by the interaction of dipoles.

Interest of researchers to the dipolar BEC \cite{Goral PRA 00}, \cite{Santos PRL 00}, \cite{Yi PRA 00} started
few years after experimental realization of BECs in vapors of alkaline atoms.
The experimental realization of dipolar BECs happened in 2005 in chromium atoms
\cite{Griesmaier PRL 05}, \cite{Lahaye Nat 07}.
In 2016 experiments show that dipolar BECs of rare-earth elements demonstrates the quantum droplets formation
\cite{Kadau Pfau Nature 16}, \cite{Ferrier-Barbut PRL 16}.
It is related to the large scale instability of dipolar BEC,
which is stabilized at smaller scales.

Quantum droplets formation in the dipolar BECs demonstrates crucial role of the quantum fluctuations
\cite{Baillie PRA 16}, \cite{Wachtler PRA 16 a2}, \cite{Bisset PRA 16}, \cite{Wachtler PRA 16 a1}, \cite{Blakie pra 16}.
Traditionally the condensate depletion is studied in terms of Bogoliubov-de Gennes theory
\cite{Lima PRA 11}, \cite{Lima PRA 12}, \cite{Blakie PRA 13},
where the depletion is presented via the Gross-Pitaevskii interaction constant.
The depletion is found in literature,
as a correction to the GP equation and the corresponding Euler equation
\cite{Ferrier-Barbut PRL 16}, \cite{Baillie PRA 16}, \cite{Bisset PRA 16},
while in our analysis is comes via additional hydrodynamic equations.
Hence, the structure of Gross-Pitaevskii equation includes the fourth order nonlinearity.
The Bogoliubov-de Gennes theory of the BEC depletion is generalized in 2012 to include the dipole-dipole interaction \cite{Lima PRA 12}.

Essential role of the quantum fluctuations in description of the quantum droplets formation is an essential part of motivation for this work.
However, obtained results give more general picture of hydrodynamics of sound waves.

General classic analysis of hydrodynamic models shows that
accurate description of the velocities of acoustic waves
requires the account of the pressure evolution equation \cite{Tokatly PRB 99}, \cite{Tokatly PRB 00},
where the acoustic waves are waves with linear spectrum $\omega^{2}=k^{2}v_{s}^{2}$,
with $\omega$ is the frequency of wave, $k$ is the wave vector, $v_{s}$ is the speed of sound.
Classical evolution of higher rank tensors leads to terms proportional to higher degrees of the wave vector $k$.

It is obtained  that there are quantum sources in equations for the higher rank tensors,
like the quantum fluctuations for BECs,
which gives contribution in the sound velocity.

\textit{Start our analysis with the microscopic Hamiltonian and present found extended set of hydrodynamic equations.}
We use the many-particle quantum hydrodynamics method,
where evolution of functions describing the collective dynamics is found
from the microscopic many-particle Schrodinger equation $\imath\hbar\partial_{t}\Psi(R,t)=\hat{H}\Psi(R,t)$ in the  coordinate representation,
where $R$ is the collection of $3N$ coordinates of $N$ particles.
Hence, collective motion is governed by the exact microscopic dynamics.
Dipolar BECs with the short-range interaction between atoms is modeled by the following Hamiltonian
\begin{equation}\label{BECTP20 Hamiltonian micro}
\hat{H}=\sum_{i=1}^{N}\biggl(\frac{\hat{\textbf{p}}^{2}_{i}}{2m_{i}}+V_{ext}(\textbf{r}_{i},t)\biggr)
+\frac{1}{2}\sum_{i,j\neq i}U_{ij}+\frac{1}{2}\sum_{i,j\neq i}U_{ij}^{d} ,\end{equation}
where $m_{i}$ is the mass of i-th particle,
$\hat{\textbf{p}}_{i}=-\imath\hbar\nabla_{i}$ is the momentum of i-th particle.
The short-range part of boson-boson interaction is presented via potential $U_{ij}=U_{SR}(\textbf{r}_{i}-\textbf{r}_{j})$.
The last term describes the long-range dipole-dipole interaction (DDI) of align dipoles \cite{Lahaye RPP 09}
$U_{ij}^{d}=\mu^{2}\frac{1-3r_{z,ij}^{2}/r_{ij}^{2}}{r_{ij}^{3}}$.
It is assumed that all dipoles are aligned parallel to the $z$-direction.

Definitely, Hamiltonian (\ref{BECTP20 Hamiltonian micro}) does not contain information about kinds of particles (bosons or fermions).
It does not include information about distribution of particles on quantum states.
However, its application to the bosons and specification of temperature
(which is a measure of the distribution on quantum states)
at the macroscopic stage of description lead to the equations for BECs dynamics.

Transition to description of the collective motion of bosons is made via introduction of the concentration
\cite{Andreev 2001}, \cite{Andreev 1912}, \cite{Andreev PRA08}, \cite{Andreev LP 19}:
\begin{equation}\label{BECTP20 concentration def b} n=\int
dR\sum_{i=1}^{N}\delta(\textbf{r}-\textbf{r}_{i})\Psi^{*}(R,t)\Psi(R,t),\end{equation}
which is the first collective variable in our model.
Other collective variables appear during the derivation.
Equation (\ref{BECTP20 concentration def b}) contains the following notations
$dR=\prod_{i=1}^{N}d\textbf{r}_{i}$ is the element of volume in $3N$ dimensional configurational space,
with $N$ is the number of bosons.

The derivation shows that
concentration (\ref{BECTP20 concentration def b}) obeys the continuity equation
\begin{equation}\label{BECTP20 cont eq via v} \partial_{t}n+\nabla\cdot (n\textbf{v})=0. \end{equation}

The velocity field $\textbf{v}$ presented in the continuity equation obeys the Euler equation,
which has the following form for bosons in the BEC state
$$mn\partial_{t}v^{\alpha} +mn(\textbf{v}\cdot\nabla)v^{\alpha}
+\partial_{\beta}T^{\alpha\beta}$$
\begin{equation}\label{BECTP20 Euler bosons BEC}
+n\partial^{\alpha}V_{ext}=-g n\partial^{\alpha}n
-n\partial^{\alpha} \Phi_{d}.\end{equation}
The Euler equation contains the short-range interaction (the first term on the right-hand side),
and the dipole-dipole interaction (the last term on the right-hand side)
in accordance with the Hamiltonian (\ref{BECTP20 Hamiltonian micro}).
The short-range interaction (SRI) contribution is obtained in the first order by the interaction radius.
Therefore, The Euler equation contains the following interaction constant
\begin{equation} \label{BECTP20 def g} g=\int d\textbf{r}U(r), \end{equation}
which is traditionally presented in the Gross-Pitaevskii equation \cite{Dalfovo RMP 99}.
The dipole-dipole interaction is presented via the macroscopic potential of dipole-dipole interaction
\begin{equation} \label{BECTP20 dd int pot def}  \Phi_{d}=\mu^{2}\int d\textbf{r}'
\frac{1}{|\textbf{r}-\textbf{r}'|^{3}}\biggl(1-3\frac{(z-z')^{2}}{|\textbf{r}-\textbf{r}'|^{2}}\biggr)
n(\textbf{r}',t), \end{equation}
since we consider it in the mean-field approximation due to its long-range nature.

The second and third terms on the right-hand side of equation (\ref{BECTP20 Euler bosons BEC})
appear as the different parts of the momentum flux $\Pi^{\alpha\beta}$.
Equations (\ref{BECTP20 cont eq via v}) and (\ref{BECTP20 Euler bosons BEC}) contain no explicit contribution of the quantum fluctuations.
If we need to find the contribution of additional effects, like the quantum fluctuations, in the hydrodynamic model,
we should extend set of hydrodynamic equations.
Therefore, we derive equations for the momentum flux second rank tensor evolution and the the third rank tensor evolution.

The left-hand side of the Euler equation contains the tensor associated with the quantum Bohm potential $T^{\alpha\beta}$.
It can be splitted on two parts $T^{\alpha\beta}=T_{0}^{\alpha\beta}+T_{qf}^{\alpha\beta}$.
The noninteracting part of the quantum Bohm potential is given by equation
\begin{equation} \label{BECTP20 Bohm tensor single part}
T_{0}^{\alpha\beta}=-\frac{\hbar^{2}}{4m^{2}}\biggl[\partial_{\alpha}\partial_{\beta}n
-\frac{\partial_{\alpha}n\cdot\partial_{\beta}n}{n}\biggr].\end{equation}
It corresponds to the Gross-Pitaevskii equation.

Analysis of further equations in the chain of hydrodynamic equations shows
that $T_{qf}^{\alpha\beta}$ has nonzero value.
Subindexes "qf" refers to the quantum fluctuations in BEC.
As it is shown below, the equation for the third rank tensor evolution
$Q_{qf}^{\alpha\beta\gamma}$ has a contribution both the short-range interaction and the DDI,
where the interaction terms are also proportional to $\hbar^{2}$.
It provides an additional contribution to Bohm potential $T_{qf}^{\alpha\beta}$.
Functions $T_{qf}^{\alpha\beta}$ and $Q_{qf}^{\alpha\beta\gamma}$ should be equal to zero at zero temperatures
(if there is no particles in the excited states).
However, the quantum terms caused by interaction leads to their nonzero value,
so some particles occupy the excited states,
we associate this contribution with the quantum fluctuations \cite{Lima PRA 11}, \cite{Lima PRA 12}.
This phenomenon is well-known in physics of quantum gases.
However, for the first time it is derived in terms of hydrodynamics model straight from microscopic quantum motion.

Equation for the nontrivial part of the momentum flux tensor $\Pi^{\alpha\beta}$,
which is the part of the quantum Bohm potential caused by the quantum fluctuations,
appears with no contribution of interaction:
\begin{equation} \label{BECTP20 eq evolution T qf}
\partial_{t}T_{qf}^{\alpha\beta} +\partial_{\gamma}(v^{\gamma}T_{qf}^{\alpha\beta})
+T_{qf}^{\alpha\gamma}\partial_{\gamma}v^{\beta}
+T_{qf}^{\beta\gamma}\partial_{\gamma}v^{\alpha}
+\partial_{\gamma}Q_{qf}^{\alpha\beta\gamma}=0.  \end{equation}
Purely quantum terms like $T_{0}^{\alpha\beta}$ cancel each other in equation (\ref{BECTP20 eq evolution T qf}).
It can be expected that $Q_{qf}^{\alpha\beta\gamma}$ is equal to zero,
but equation for its evolution shows that it is not equal to zero even for the BECs.

Equation for the evolution of quantum-thermal part of the third rank tensor is:
$$\partial_{t}Q_{qf}^{\alpha\beta\gamma} +\partial_{\delta}(v^{\delta}Q_{qf}^{\alpha\beta\gamma})
+Q_{qf}^{\alpha\gamma\delta}\partial_{\delta}v^{\beta}
+Q_{qf}^{\beta\gamma\delta}\partial_{\delta}v^{\alpha}
+Q_{qf}^{\alpha\beta\delta}\partial_{\delta}v^{\gamma}$$
$$+\partial_{\delta}P_{qf}^{\alpha\beta\gamma\delta}
=\frac{\hbar^{2}}{4m^{3}} n\biggl(g_{2}I_{0}^{\alpha\beta\gamma\delta}\partial^{\delta}n
+\partial^{\alpha}\partial^{\beta}\partial^{\gamma}\Phi_{d}\biggr)$$
\begin{equation} \label{BECTP20 eq evolution Q qf}
+\frac{1}{mn}(T_{qf}^{\alpha\beta}\partial^{\delta}T_{qf}^{\gamma\delta}
+T_{qf}^{\alpha\gamma}\partial^{\delta}T_{qf}^{\beta\delta}
+T_{qf}^{\beta\gamma}\partial^{\delta}T_{qf}^{\alpha\delta}),  \end{equation}
where
\begin{equation} \label{BECTP20 I 4} I_{0}^{\alpha\beta\gamma\delta}=\delta^{\alpha\beta}\delta^{\gamma\delta} +\delta^{\alpha\gamma}\delta^{\beta\delta}+\delta^{\alpha\delta}\delta^{\beta\gamma}. \end{equation}
Equation (\ref{BECTP20 eq evolution Q qf}) is the reduction of the third rank tensor evolution equation for the BECs.
Equation (\ref{BECTP20 eq evolution Q qf}) contains the second interaction constant for the short-range interaction
\begin{equation} \label{BECTP20 def g 2} g_{2}=\frac{2}{3}\int d\textbf{r} U''(r). \end{equation}
This interaction constant is proportional to the zeroth order moment of the second derivative of the potential of the short-range interaction,
while the first interaction constant is the zeroth order moment of the potential of the short-range interaction.
The second interaction constant appears in the first order by the interaction radius like the interaction in the Gross-Pitaevskii,
but for the evolution of physical function of higher tensor rank.

For truncation of obtained set of equations we assume
that $P_{qf}^{\alpha\beta\gamma\delta}=0$.

The developed model contains the unknown parameter $g_{2}$.
This is a parameter independent from interaction constant $g$.
It would be methodologically incorrect to give an estimation of $g_{2}$ via $g$.
Moreover, the second interaction constant $g_{2}$ is not related to interaction constants introduced in
Refs. \cite{Andreev 2001}, \cite{Andreev 1912}, \cite{Andreev PRA08}, \cite{Andreev LP 19}, \cite{Rosanov},
\cite{Braaten},
where the additional constants appear in the Euler equation
at more detailed description of the force field in the third order by the interaction radius
$g_{T}\sim\int d\textbf{r} r^{2}U(r)$.
Some arbitrary values of interaction constant $g_{2}$ are used below for estimation of its contribution in the spectrum.

It can be useful to mention that
the account of the short-range interaction in the Euler equation constant leads
to additional interaction constant,
which is the second order moment of the potential of the short-range interaction \cite{Andreev PRA08}.
Hence, extension of the BEC models beyond the Gross-Pitaevskii approximation gives additional characteristics of potential
which can be measured so the potential can be found with some accuracy.

Obtained structure of hydrodynamic equations is correct for the fermions as well.
However, the first-order by the interaction radius is equal to zero.
However, terms proportional to $g_{2}$ should appear in the next orders.
Moreover, the form of dipole-dipole interaction is the same for fermions.

Appearance of the interaction constants (\ref{BECTP20 def g}) and (\ref{BECTP20 def g 2}) is not related to the scattering problem
or the application of Bohm approximation.
They are direct consequence of the small radius nature of the interaction given by potential $U(r)$.
They appear before we make any judgement about strength of interaction,
while explicit form of interaction terms proportional to derivative of concentration square $\nabla n^{2}$
are consequences of the weak interaction limit.

\textit{Calculate spectrum of bulk collective excitations.}
It appears as the generalization of the well-known Bogoliubov spectrum,
where the generalization is caused by the quantum fluctuations.
Consider small amplitude perturbations of the equilibrium state
while the equilibrium state is described by the constant nonzero concentration $n_{0b}$,
zero value velocity $\textbf{v}_{0b}=0$,
and the zero value quantum Bohm tensor $T_{0}^{\alpha\beta}=0$.
Small perturbation of each function is considered as plane waves propagating parallel to the $x$-direction,
for instance for concentration
$\delta n=Ne^{-\imath \omega t+\imath k_{x}x+\imath k_{z}z}$,
where $N$ is the amplitude of perturbation.

First, we stress our attention on the quantum fluctuations caused by the short-range interaction.
Consider BECs of vapors of alkaline atoms, where the dipole moments gives no noticeable contribution.
Change the interaction constant to zero value by the Feshbach resonance $g=0$
and drop the contribution of the noninteracting part of the quantum Bohm potential $T_{0}^{\alpha\beta}$.
Equation (\ref{BECTP20 eq evolution Q qf}) simplifies to
\begin{equation} \label{BECTP20 Q eq simplified for qf SRI lin}
\partial_{t}\delta Q^{\alpha\beta\gamma}=\frac{\hbar^{2}}{4m^{3}}g_{2}I_{0}^{\alpha\beta\gamma\delta}n_{0}\partial^{\delta}\delta n. \end{equation}
The Euler equation (\ref{BECTP20 Euler bosons BEC}) has zero right-hand side in this limit.
Therefore,
the spectrum of bulk excitation shows linear dependence of the frequency on the wave vector
\begin{equation} \label{BECTP20 spectrum qf SRI}
\omega^{2}=\frac{\sqrt{3(-g_{2})n_{0}}\hbar}{2m^{2}} k^{2}.  \end{equation}
The second interaction constant $g_{2}$ defines the speed of sound.
Moreover, it should be negative to get a stable solution.

However, for repulsive interaction $U>0$,
we normally have $U''>0$.
It gives positive second interaction constant.
Hence, equation (\ref{BECTP20 spectrum qf SRI}) shows an instability.

Next, we present the dispersion equation in general regime.
$$\omega^{4}-\biggl[\frac{n_{0}}{m}\biggl(g+\mu^{2}(\cos^{2}\theta-1/3)\biggr)k^{2}+\frac{\hbar^{2}k^{4}}{4m^{2}}\biggr]\omega^{2}$$
\begin{equation} \label{BECTP20 disp eq with qf}
-\frac{\hbar^{2}k^{4}n_{0}}{m^{3}}\biggl(\mu^{2}k^{2}(\cos^{2}\theta-1/3)-3g_{2}\biggr)=0,  \end{equation}
where $k^{2}=k_{x}^{2}+k_{z}^{2}$.

First two terms in equation (\ref{BECTP20 disp eq with qf}) are the source of the traditional Bogoliubov spectrum of dipolar BECs.
The last term presents the quantum fluctuations.

Equation (\ref{BECTP20 disp eq with qf}) has two solutions.
One solution is the generalization Bogoliubov spectrum.
The second solution is novel solution
which exists
if the last term in equation (\ref{BECTP20 disp eq with qf}) is positive.

In the second term,
the repulsive SRI combines with the repulsive DDI to increase the coefficient.
However, the last term demonstrates different relation between the SRI and the DDI.
There is the competition of these interactions if both interactions are repulsive interactions.
Reason for such difference is in the following.
The SRI is proportional to the first derivative of concentration in both equations containing the SRI.
It is the Euler equation and the equation for evolution of the third rank tensor $Q^{\alpha\beta\gamma}_{qf}$.
The DDI shows different picture since it is the long-range interaction.
The DDI is always proportional to the macroscopic potential (\ref{BECTP20 dd int pot def}).
However, the Euler equation contains the first derivative of the potential,
while the third rank tensor evolution equation includes the third derivative of the potential.
At the transition to the plane waves two additional derivatives gives minus.
In the SRI term extra derivatives are hidden is the second interaction constant.

\begin{figure}
\includegraphics[width=8cm,angle=0]{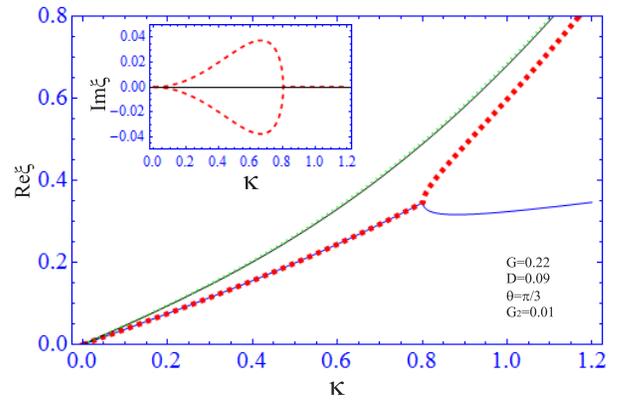}
\caption{\label{BECTP20 Fig 01}
Real and imaginary parts of frequency in the regime of attractive DDI.
Dotted curve shows the spectrum at the zero dipole moment and the zero quantum fluctuations.
Thin continuous curve gives spectrum of dipolar BEC at the zero quantum fluctuations.
Thick continuous curve and thick dashed curve show the spectrum of dipolar BEC under influence of the quantum fluctuations.}
\end{figure}

\begin{figure}
\includegraphics[width=8cm,angle=0]{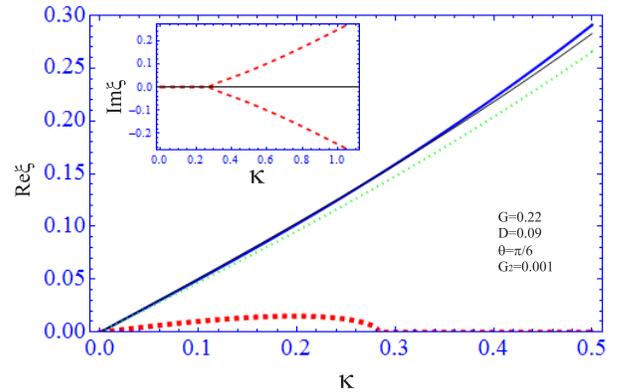}
\caption{\label{BECTP20 Fig 02}
Real and imaginary parts of frequency in the regime of repulsive DDI,
where the dipolar quantum fluctuations dominate over the quantum fluctuations caused by the SRI.
The curves description is the same as in Fig. \ref{BECTP20 Fig 01}.}
\end{figure}

\begin{figure}
\includegraphics[width=8cm,angle=0]{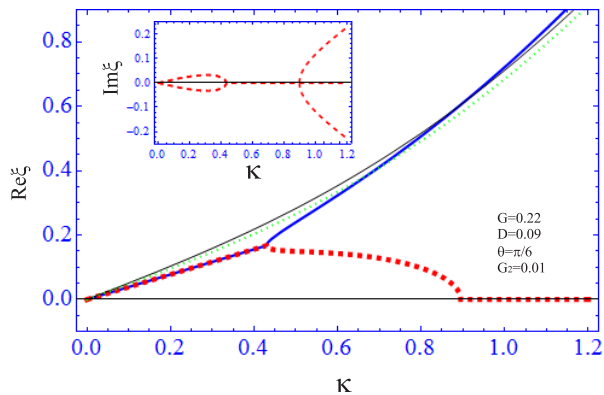}
\caption{\label{BECTP20 Fig 03}
Real and imaginary parts of frequency in the regime of repulsive DDI
for the stronger part of quantum fluctuations caused by the SRI, in compare with Fig. \ref{BECTP20 Fig 02}.
The curves description is the same as in Fig. \ref{BECTP20 Fig 01}.}
\end{figure}

Strong role of quantum fluctuations is demonstrated in $^{164}$Dy BEC,
where the quantum fluctuations cause the quantum droplets formation \cite{Kadau Pfau Nature 16}.
Atoms of $^{164}$Dy have relatively large magnetic moment $\mu=10\mu_{B}$, with $\mu_{B}$ is the Bohr magneton.
SRI in this is characterized by the following value of the scattering length $a=70a_{B}$, with $a_{B}$ is the Bohr radius, and $g=4\pi\hbar^{2}a/m$.


We study the spectrum and mechanisms for instabilities for the uniform dipolar BEC.
The equilibrium concentration is chosen to be $n_{0}=10^{14}$ cm$^{-3}$,
which corresponds to the average concentrations of the trapped BECs in existing experiments.
Numerical analysis of spectra is made in terms of the dimensionless parameters
for the wave vector $\kappa=k/n_{0}^{1/3}$,
frequency $\xi=m\omega/\hbar n_{0}^{2/3}$,
dipole-dipole interaction strength $D=m\mu^{2}n_{0}^{1/3}/\hbar^{2}$,
and the short-range interaction constants $g=G\hbar^{2}/mn_{0}^{1/3}$, $g_{2}=G_{2}\hbar^{2}n_{0}^{1/3}/m$.
The first interaction constant can be represented via the scattering length $G=4\pi an_{0}^{1/3}$.
We have the following values for the $^{164}$Dy BEC:
$G=0.22$ for $a=70a_{B}$,
$D=0.09$ for $a_{dd}=131a_{B}$ corresponding to $\mu=10\mu_{B}$, since $a_{dd}\equiv m\mu^{2}/3\hbar^{2}$, and
$D=3a_{dd}n_{0}^{1/3}$.

Unstable regimes given by equation (\ref{BECTP20 disp eq with qf}) are demonstrated in Figs.
\ref{BECTP20 Fig 01}, \ref{BECTP20 Fig 02}, \ref{BECTP20 Fig 03}.

The well-known instability causing collapse can appear if we have large DDI in the attractive regime.
However, we have $G=0.22$, $D=0.09$, and $D(\cos^{2}\theta-1/3)\mid_{\theta=\pi/3}=-0.0075\ll G$.
Hence, the Bogoliubov spectrum is stable.
There is no large competition between $G$ and $D$ in the traditional spectrum for used parameters.

Another example of competition in presented by the last term in equation (\ref{BECTP20 disp eq with qf})
which is caused by the quantum fluctuations.

The stability of the spectrum depends on the sign of the determinant $\Delta$ of dispersion equation (\ref{BECTP20 disp eq with qf}).
If the last term in equation (\ref{BECTP20 disp eq with qf}) is negative it makes the determinant $\Delta$ positive.
However, the square of frequency for the second solution of equation (\ref{BECTP20 disp eq with qf}) is negative in this case
(the first solution is associated with the Bogoliubov spectrum).
It gives condition for the instability demonstrated in Fig. \ref{BECTP20 Fig 01} 
and the small wave vector area of Fig. \ref{BECTP20 Fig 03}.

If the last term in equation (\ref{BECTP20 disp eq with qf}) is positive we can have positive square of frequency for the second solution,
but determinant $\Delta$ can become negative for the large quantum fluctuations presented by the last term in equation (\ref{BECTP20 disp eq with qf}).
Therefore, there is the mechanism for the instability.

The dipolar part of quantum fluctuations is not a constant,
but it suppressed by the small dimensionless wave vectors $\kappa<1$.
However, we expect that $G_{2}\ll G$.
Hence, the dipolar contribution can overcome $G_{2}$ at $\kappa<1$.
For repulsive interaction we have $G_{2}>0$.
It means that the attractive DDI increases the contribution of the SRI.
We have competition of two terms
if the DDI is repulsive.
For instance, if $D=0.09$, $\theta=\pi/6$, $(\cos^{2}\theta-1/3)=5/12$, and $G_{2}=0.01$.
Hence, the critical wave vector is $\kappa=0.89$
(see corresponding point in Fig. \ref{BECTP20 Fig 03}).


\emph{To conclude} we mention that
the extended hydrodynamic model of dipolar BECs has been developed
to give a purely hydrodynamic description of quantum fluctuations.
It has been found that
the short-range interaction proportional to the zeroth moment of the second derivatives of the interaction potential
and the third derivative of the macroscopic potential of dipole-dipole interaction are responsible for the quantum fluctuation appearance.
These terms are also proportional to the square of the Planck constant.
These terms are presented in the third rank tensor evolution equation,
while the second rank tensor (superposition of the pressure and the quantum Bohm potential) evolution equation has no contribution of interaction.
Therefore, found extended hydrodynamics consists of four equations for material fields of different tensor ranks:
the continuity equation for the concentration,
the Euler equation for the velocity vector field,
the pressure second rank tensor evolution equation
(the quantum pressure or the quantum Bohm potential caused by the quantum fluctuations)
and the evolution equation for the third rank tensor.

The quantum fluctuations cause the depletion of BEC,
so some excited states are occupied.
The contribution of excited states has been mainly modeled by the nontrivial part of the quantum Bohm potential and the third rank tensor.
Developed model has been applied to study the bulk excitations in the uniform BECs.
Hance, a generalization of the Bogoliubov spectrum has been obtained.

It has been obtained that
the quantum fluctuations can cause the long-wavelength instability.
Moreover, in the stability regime there are two wave solutions,
where the second wave is caused by the quantum fluctuations.
The second wave can go unstable at the small wavelengths.

\textit{Acknowledgements}
Work is supported by the Russian Foundation for Basic Research (grant no. 20-02-00476).

\newpage

\section{Suplementerly materials}

\subsection{Definitions of basic hydrodynamic variables}

After derivation of the continuity equation (\ref{BECTP20 cont eq via v})
for concentration (\ref{BECTP20 concentration def b})
from the Schrodinger equation with Hamiltonian (\ref{BECTP20 Hamiltonian micro}),
the current appears as the following integral of the wave function
$$\textbf{v}=\frac{\textbf{j}}{n}
=\frac{1}{n}\int dR\sum_{i=1}^{N}\delta(\textbf{r}-\textbf{r}_{i})\times$$
\begin{equation}\label{BECTP20 j def}
\times\frac{1}{2m_{i}}(\Psi^{*}(R,t)\hat{\textbf{p}}_{i}\Psi(R,t)+c.c.),\end{equation}
with $c.c.$ is the complex conjugation.

Definition of current (\ref{BECTP20 j def}) allows to derive the Euler equation for the current (momentum density) evolution
\begin{equation} \label{BECTP20 Euler eq 1 via j}
\partial_{t}j^{\alpha}+\partial_{\beta}\Pi^{\alpha\beta}
=-\frac{1}{m}n\partial_{\alpha}V_{ext}+\frac{1}{m}F^{\alpha}_{int}, \end{equation}
where
$$\Pi^{\alpha\beta}=\int dR\sum_{i=1}^{N}\delta(\textbf{r}-\textbf{r}_{i}) \frac{1}{4m^{2}}
[\Psi^{*}(R,t)\hat{p}_{i}^{\alpha}\hat{p}_{i}^{\beta}\Psi(R,t)$$
\begin{equation} \label{BECTP20 Pi def} +\hat{p}_{i}^{\alpha *}\Psi^{*}(R,t)\hat{p}_{i}^{\beta}\Psi(R,t)+c.c.] \end{equation}
is the momentum flux,
and
\begin{equation} \label{BECTP20 F alpha def via n2}
F^{\alpha}_{int}=-\int (\partial^{\alpha}U(\textbf{r}-\textbf{r}'))
n_{2}(\textbf{r},\textbf{r}',t)d\textbf{r}', \end{equation}
with the two-particle concentration
$$n_{2}(\textbf{r},\textbf{r}',t)$$
\begin{equation} \label{BECTP20 n2 def} =\int
dR\sum_{i,j=1,j\neq i}^{N}\delta(\textbf{r}-\textbf{r}_{i})\delta(\textbf{r}'-\textbf{r}_{j})\Psi^{*}(R,t)\Psi(R,t) .\end{equation}

It gives the general structure of the Euler equation and the definition of the momentum flux.

\subsection{General structure of equation for the second order tensor}

Extending the set of hydrodynamic equations we can derive the equation for the momentum flux evolution.
Consider the time evolution of the momentum flux (\ref{BECTP20 Pi def}) using the Schrodinger equation with Hamiltonian (\ref{BECTP20 Hamiltonian micro})
and derive the momentum flux evolution equation
$$\partial_{t}\Pi^{\alpha\beta}+\partial_{\gamma}M^{\alpha\beta\gamma}
=-\frac{1}{m}j^{\beta}\partial_{\alpha}V_{ext}$$
\begin{equation} \label{BECTP20 eq for Pi alpha beta} -\frac{1}{m}j^{\alpha}\partial_{\beta}V_{ext}
+\frac{1}{m}(F^{\alpha\beta}+F^{\beta\alpha}), \end{equation}
where
$\Pi^{\alpha\beta}=\Pi_{n}^{\alpha\beta}+\Pi_{b}^{\alpha\beta}$,
\begin{equation} \label{BECTP20 F alpha beta def} F^{\alpha\beta}=-\int[\partial^{\alpha}U(\textbf{r}-\textbf{r}')]j_{2}^{\beta}(\textbf{r},\textbf{r}',t)d\textbf{r}',\end{equation}
$$M^{\alpha\beta\gamma}=\int dR\sum_{i=1}^{N}\delta(\textbf{r}-\textbf{r}_{i}) \frac{1}{8m_{i}^{3}}\biggl[\Psi^{*}(R,t)\hat{p}_{i}^{\alpha}\hat{p}_{i}^{\beta}\hat{p}_{i}^{\gamma}\Psi(R,t)$$
$$+\hat{p}_{i}^{\alpha *}\Psi^{*}(R,t)\hat{p}_{i}^{\beta}\hat{p}_{i}^{\gamma}\Psi(R,t)
+\hat{p}_{i}^{\alpha *}\hat{p}_{i}^{\gamma *}\Psi^{*}(R,t)\hat{p}_{i}^{\beta}\Psi(R,t)$$
\begin{equation} \label{BECTP20 M alpha beta gamma def}
+\hat{p}_{i}^{\gamma *}\Psi^{*}(R,t)\hat{p}_{i}^{\alpha}\hat{p}_{i}^{\beta}\Psi(R,t)+c.c.\biggr], \end{equation}
and
$$\textbf{j}_{2}(\textbf{r},\textbf{r}',t)=\int
dR\sum_{i,j\neq i}\delta(\textbf{r}-\textbf{r}_{i})\delta(\textbf{r}'-\textbf{r}_{j})\times$$
\begin{equation} \label{BECTP20 j 2 def}
\times\frac{1}{2m_{i}}(\Psi^{*}(R,t)\hat{\textbf{p}}_{i}\Psi(R,t)+c.c.) .\end{equation}
If quantum correlations are dropped function $j_{2}^{\alpha}(\textbf{r},\textbf{r}',t)$
splits on product of the current $j^{\alpha}(\textbf{r},t)$ and the concentration $n(\textbf{r}',t)$.
Tensor $M^{\alpha\beta\gamma}$ (\ref{BECTP20 M alpha beta gamma def}) is the flux of the momentum flux.
Interaction in the momentum flux evolution equation (\ref{BECTP20 eq for Pi alpha beta}) is presented
by symmetrized combinations of tensors $F^{\alpha\beta}$,
which is the flux or current of force.

The pressure is the average of the square of the thermal velocity,
when tensor $Q^{\alpha\beta\gamma}$ is the average of the product of three projections of the thermal velocity.
For the BEC we have $p_{B}^{\alpha\beta}=0$, and $Q_{B}^{\alpha\beta\gamma}=0$.
Function $Q_{qf}^{\alpha\beta\gamma}\equiv L_{B}^{\alpha\beta\gamma}$ is the thermal-quantum term,
where both contributions are intertwine together
(the general structure of $L^{\alpha\beta\gamma}$ is introduced in Ref. \cite{Andreev 2001}).
The notion "thermal" refers to the presence of particles in the excited states,
while nature of the excitation can be arbitrary.
In our case, the reason of excitation is the interaction existing in third rank tensor evolution equation.

The equation for evolution of the third rank tensor is derived for $M^{\alpha\beta\gamma}$ (\ref{BECTP20 M alpha beta gamma def}).
It contains some contribution of the interaction,
similarly to the right-hand side of equation (\ref{BECTP20 eq for Pi alpha beta}).
The structure of equation changes after the introduction of the velocity field $\textbf{v}$
(compare for instance equations (\ref{BECTP20 eq evolution T qf}) and (\ref{BECTP20 eq for Pi alpha beta})).
The contribution of interaction partially cancels via the time derivatives of the velocity field given by the Euler equation (\ref{BECTP20 Euler bosons BEC}).

Methods of calculation of the terms containing the short-range interaction are presented in
Refs. \cite{Andreev 2001}, \cite{Andreev 1912}, \cite{Andreev PRA08}.
Refs. \cite{Andreev 2001}, \cite{Andreev 1912} are focused on ultracold fermions,
but methodology is the same.

\begin{figure}
\includegraphics[width=8cm,angle=0]{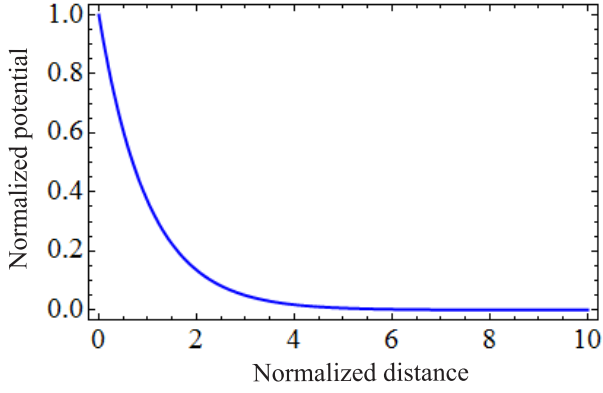}
\caption{\label{BECTP20 Fig 00}
Normalized potential of the short-range interaction is plotted to show that it second derivative is positive $U''>0$.}
\end{figure}

\begin{figure}
\includegraphics[width=8cm,angle=0]{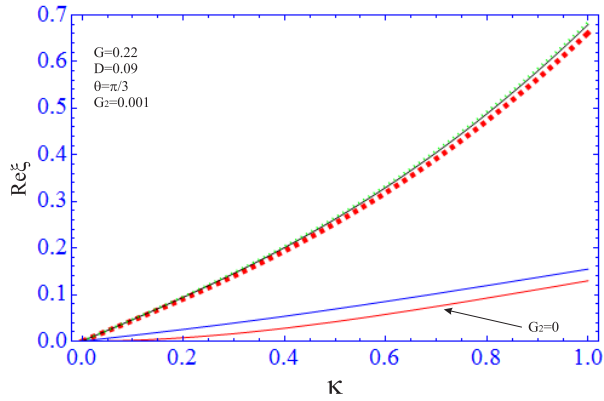}
\caption{\label{BECTP20 Fig 04}
Regime of the small quantum fluctuations related to the SRI is presented for the attractive DDI.
Regime of quantum fluctuations caused purely by dipoles (limit of $g_{2}=0$) is included
by the lower continuous red curve.
Dotted curve shows the spectrum at the zero dipole moment and the zero quantum fluctuations.
Thin continuous curve gives spectrum of dipolar BEC at the zero quantum fluctuations.
Thick continuous curve and thick dashed curve show the spectrum of dipolar BEC under influence of the quantum fluctuations.
}
\end{figure}



\subsection{Equation for evolution of the fourth rank tensor}

To understand the approximation given by equations
(\ref{BECTP20 cont eq via v}), (\ref{BECTP20 Euler bosons BEC}), (\ref{BECTP20 eq evolution T qf}), and (\ref{BECTP20 eq evolution Q qf})
we need to consider equations for the higher rank tensors.

Equation for the quantum-thermal part of the fourth rank tensor
$$\partial_{t}P_{qf}^{\alpha\beta\gamma\delta}
+\partial_{\nu}(v^{\nu}P_{qf}^{\alpha\beta\gamma\delta})
+P_{qf}^{\beta\gamma\delta\nu}\partial_{\nu}v^{\alpha}
+P_{qf}^{\alpha\gamma\delta\nu}\partial_{\nu}v^{\beta}$$
$$+P_{qf}^{\alpha\beta\delta\nu}\partial_{\nu}v^{\gamma}
+P_{qf}^{\alpha\beta\gamma\nu}\partial_{\nu}v^{\delta}
=\frac{1}{mn}\biggl[Q_{qf}^{\beta\gamma\delta}\partial^{\nu}T^{\alpha\nu}$$
\begin{equation} \label{BECTP20 eq evolution P 4 qf}
+Q_{qf}^{\alpha\gamma\delta}\partial^{\nu}T^{\beta\nu}
+Q_{qf}^{\alpha\beta\delta}\partial^{\nu}T^{\gamma\nu}
+Q_{qf}^{\alpha\beta\gamma}\partial^{\nu}T^{\delta\nu}\biggr]  \end{equation}
is also obtained.
There is no interaction contribution in this equation.
We have general tendency that equations for evolution of the even rank tensors have no contribution of interaction.
However, equations for evolution of the odd rank tensors have contribution of interaction.
This interaction has nonzero contribution in the first order by the interaction radius.
New interaction constants appear in each equation.
To illustrate the last statement we present a part of the fifth rank tensor evolution equation.

The divergence of the quantum-thermal part of the fifth rank tensor is dropped.
Its thermal part has zero value in the equilibrium state, so it can be used as an equation of state.
The mixed quantum-thermal part $R^{\alpha\beta\gamma\delta\mu}$ is also assumed to be equal to zero.
However, there is $R_{qf}^{\alpha\beta\gamma\delta\mu}$ caused by the quantum fluctuations of higher order,
but being obtained in the first order by the interaction radius.
For the SRI, the quantum fluctuations is proportional to $n\nabla n$, and the the third interaction constant $g_{3}\sim \int d\textbf{r} U^{(4)}(r)$,
where $U^{(4)}(r)$ is the fourth derivative of the SRI potential.
For the DDI we have that the time derivative of $R_{qf}^{\alpha\beta\gamma\delta\mu}$
is caused by the fifth space derivative of the macroscopic potential of DDI $\Phi_{d}$.

So, the evolution each tensor of uneven rank $A^{2n-1}_{qf}$ gives
the contribution of higher order quantum fluctuations via new interaction constant
$g_{n}\sim \int d\textbf{r} \frac{d^{2n-2}U(r)}{dr^{2n-2}}$ for the SRI.
Its evolution is proportional to the space derivative of $2n-1$ order of $\Phi_{d}$ for the DDI.

\subsection{Linearized hydrodynamic equations}

The linear approximation of the hydrodynamic equations
(\ref{BECTP20 cont eq via v})-(\ref{BECTP20 eq evolution Q qf}) has the following form:
\begin{equation} \label{BECTP20 cont lin} \omega\delta n=n_{0}(k_{x}\delta v^{x}+k_{z}\delta v^{z}), \end{equation}
$$\omega mn_{0}\delta v^{x}-k_{x} \delta T_{qf}^{xx} -k_{z} \delta T_{qf}^{xz}$$
\begin{equation} \label{BECTP20 Euler lin x}
-\frac{\hbar^{2}k^{2}}{4m}k_{x}\delta n
=gn_{0}k_{x}\delta n+n_{0}k_{x}\delta\Phi_{d},\end{equation}
$$\omega mn_{0}\delta v^{z}-k_{x} \delta T_{qf}^{xz} -k_{z} \delta T_{qf}^{zz}$$
\begin{equation} \label{BECTP20 Euler lin z}
-\frac{\hbar^{2}k^{2}}{4m}k_{z}\delta n
=gn_{0}k_{z}\delta n+n_{0}k_{z}\delta\Phi_{d},\end{equation}
\begin{equation} \label{BECTP20 pressure evol lin xx}
\omega \delta T_{qf}^{xx}=k_{x}\delta Q_{qf}^{xxx}+k_{z}\delta Q_{qf}^{xxz},\end{equation}
\begin{equation} \label{BECTP20 pressure evol lin zz}
\omega \delta T_{qf}^{zz}=k_{x}\delta Q_{qf}^{zzx}+k_{z}\delta Q_{qf}^{zzz},\end{equation}
\begin{equation} \label{BECTP20 pressure evol lin xz}
\omega \delta T_{qf}^{xz}=k_{x}\delta Q_{qf}^{xxz}+k_{z}\delta Q_{qf}^{xzz},\end{equation}
\begin{equation} \label{BECTP20 Q evol lin} \omega \delta Q^{xxx}=
\frac{\hbar^{2}}{4m^{3}} n_{0}\biggl(k_{x}^{3}\delta\Phi_{d}
-3k_{x} g_{2}  \delta n \biggr), \end{equation}
\begin{equation} \label{BECTP20 Q evol lin} \omega \delta Q^{zzz}=
\frac{\hbar^{2}}{4m^{3}} n_{0}\biggl(k_{z}^{3}\delta\Phi_{d}
-3k_{z} g_{2}  \delta n \biggr), \end{equation}
\begin{equation} \label{BECTP20 Q evol lin} \omega \delta Q^{xxz}=
\frac{\hbar^{2}}{4m^{3}} n_{0}k_{z}\biggl(k_{x}^{2}\delta\Phi_{d}
-g_{2}  \delta n \biggr), \end{equation}
\begin{equation} \label{BECTP20 Q evol lin} \omega \delta Q^{xzz}=
\frac{\hbar^{2}}{4m^{3}} n_{0}k_{x}\biggl(k_{z}^{2}\delta\Phi_{d}
-g_{2}  \delta n \biggr), \end{equation}
where $I_{0}^{xxxx}=I_{0}^{zzzz}=3$,
$I_{0}^{xxzz}=1$, $I_{0}^{xzzz}=0$.

Linearized potential of dipole-dipole interaction is \cite{Lahaye RPP 09}
\begin{equation} \label{BECTP20 dd int pot lin}\delta\Phi_{d}=\mu^{2}(\cos^{2}\theta-1/3)\delta n.\end{equation}

Equations (\ref{BECTP20 cont lin})-(\ref{BECTP20 dd int pot lin}) are used to obtain spectra (\ref{BECTP20 spectrum qf SRI})-(\ref{BECTP20 disp eq with qf}).

\subsection{Signature of the second interaction constant}

It is used in the text that
for the repulsive interaction $g_{2}>0$.
Fig. (\ref{BECTP20 Fig 00}) demonstrates the simple example of repulsive potential.
It shows that $U''>0$, hence $g_{2}>0$.

\subsection{Dimensionless form of dispersion equation}

Present the dispersion equation (\ref{BECTP20 disp eq with qf}) in dimensionless form
for zero dipole contribution
\begin{equation} \label{BECTP20} \xi^{4} -[G \kappa^{2}+0.25 \kappa^{4}]\xi^{2} +3\kappa^{4} G_{2}=0,\end{equation}
where
dimensionless wave vector $\kappa=k/n_{0}^{1/3}$,
dimensionless frequency $\xi=m\omega/\hbar n_{0}^{2/3}$,
dimensionless interaction constants $g=G\hbar^{2}/mn_{0}^{1/3}$, and $g_{2}=G_{2}\hbar^{2}n_{0}^{1/3}/m$, or $G=4\pi an_{0}^{1/3}$.

Consider dimensionless form of equation (\ref{BECTP20 disp eq with qf})
for dipolar BECs
$$\xi^{4} -[G \kappa^{2}+D(\cos^{2}\theta-1/3)\kappa^{2}+0.25 \kappa^{4}]\xi^{2}$$
\begin{equation} \label{BECTP20} -[D(\cos^{2}\theta-1/3)\kappa^{2}-3 G_{2}]\kappa^{4}=0,\end{equation}
where
$D=m\mu^{2}n_{0}^{1/3}/\hbar^{2}$.

\subsection{Spectrum: small quantum fluctuation limit}

No instability appears
if quantum fluctuations are dominated by the attractive DDI.
However, the second stable low frequency wave solution appears in this regime,
as it is shown by the two lower lines in Fig. (\ref{BECTP20 Fig 04}).

\end{document}